\pdfoutput=1
\documentclass[aps,twocolumn, floatfix, prl]{revtex4}
\usepackage{graphicx}
\DeclareGraphicsExtensions{.pdf}
\usepackage{amsmath,amssymb,bbold,bm}
\usepackage{float}
\usepackage{epstopdf}

\newcommand{\bk}{{\bm k}}
\newcommand{\br}{{\bm r}}

\newcommand{\bB}{{\bm B}}
\newcommand{\bE}{{\bm E}}
\newcommand{\bb}{{\bm b}}
\newcommand{\bA}{{\bm A}}

\newcommand{\bs}{{\bm s}}
\newcommand{\bj}{{\bm j}}

\newcommand{\cT}{{\cal T}}
\newcommand{\cP}{{\cal P}}

\begin{document}

\title{Electromagnetic Response of Weyl Semimetals}

\author{M.M. Vazifeh}
\author{M. Franz}
\affiliation{Department of Physics and Astronomy, University of
British Columbia, Vancouver, BC, Canada V6T 1Z1}

\begin{abstract}
It has been suggested recently, based on subtle field-theoretical considerations, that the electromagnetic response of Weyl semimetals and the closely related Weyl insulators can be characterized by an axion term $\theta {\bm E}\cdot \bB$ with space and time dependent axion angle $\theta(\br,t)$. Here we construct a minimal  lattice model of the Weyl medium and study its electromagnetic response by a combination of analytical and numerical techniques. We confirm the existence of the anomalous Hall effect expected on the basis of the field theory treatment. We find, contrary to the latter, that chiral magnetic effect (that is, ground-state charge current induced by the applied magnetic field) is absent in both the semimetal and the insulator phase.   We elucidate the reasons for this discrepancy.
\end{abstract}

\date{\today}

\maketitle

\begin{figure}[t]
\includegraphics[width = 8cm]{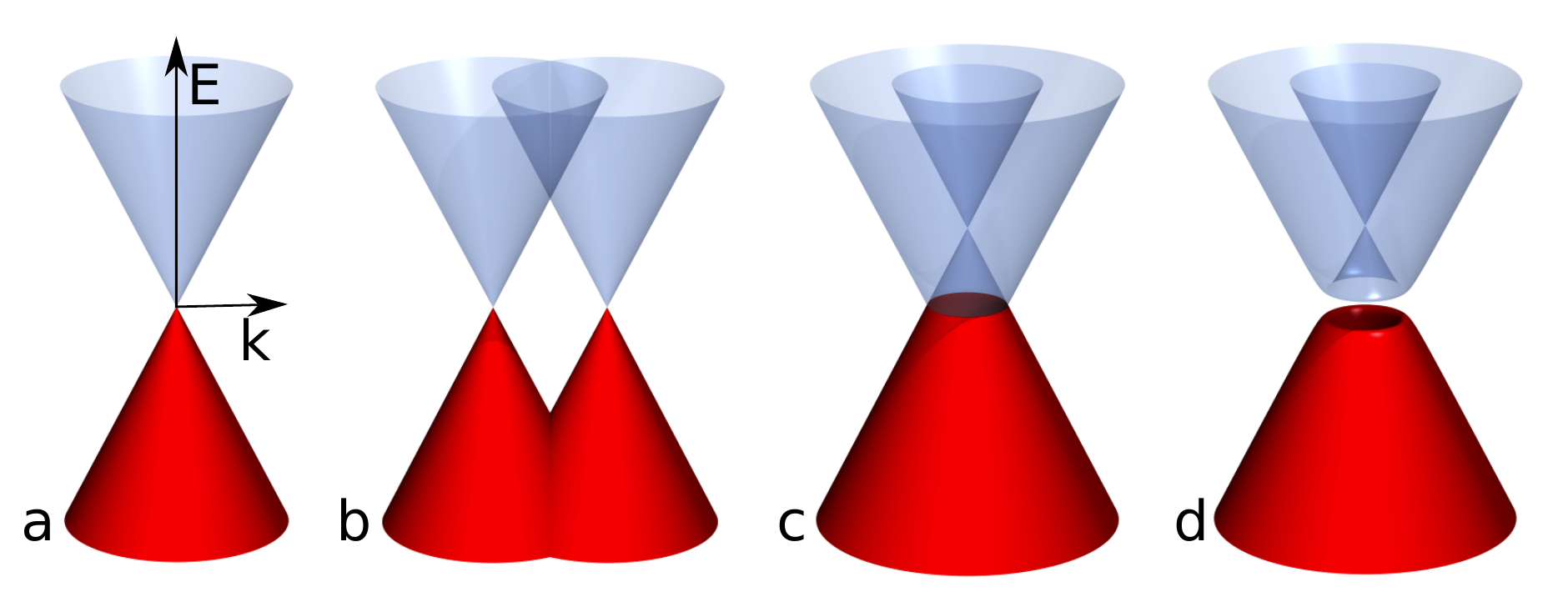}
\caption{Low energy spectra in Dirac and Weyl semimetals. a) Doubly degenerate massless Dirac cone at the transition from a TI to a band insulator. Weyl semimetals with the individual cones shifted in b) momenta and c) energy.  Panel d) illustrates the Weyl insulator which can arise when the excitonic instability gaps out the spectrum indicated in c). In all panels two components of the 3D crystal momentum $\bk$ are shown. 
}\label{fig1}
\end{figure}
When a three-dimensional  topological insulator (TI) \cite{mooreN,hasan_rev,qi_rev} undergoes a phase transition into an ordinary band insulator, its low-energy electronic spectrum at the critical point consists of an odd number of 3D massless Dirac points. Such 3D Dirac points have been experimentally observed in TlBi(S$_{1-x}$Se$_x$)$_2$ crystals \cite{xu1} and in (Bi$_{1-x}$In$_x$)$_2$Se$_2$ films \cite{brahlek1}. 
In the presence of the time reversal ($\cT$) and inversion ($\cP$) symmetries  the Dirac points are doubly degenerate and occur at high-symmetry positions in the Brillouin zone. When $\cT$ or $\cP$ is broken, however, each Dirac point can split into a pair of `Weyl points' separated from one another in momentum $\bk$ or energy $E$, as illustrated in Fig.\ \ref{fig1}. The resulting Weyl semimetal constitutes a new phase of topological quantum matter \cite{murakami1,weyl4,witcak1,weyl1,weyl2,weyl3,halasz1,cho1,liu1} with a number of fascinating physical properties including protected surface states and unusual electromagnetic response. 

The low energy theory of  an isolated  Weyl point is given by the Hamiltonian
\begin{equation}\label{hw}     
h_W(\bk)=b_0 + v{\bm \sigma}\cdot(\bk-\bb),
\end{equation}
where $v$ is the characteristic velocity, ${\bm \sigma}$ a vector of the Pauli matrices, $b_0$ and $\bb$ denote the shift in energy and momentum, respectively. Because all three Pauli matrices are used up in $h_W(\bk)$, small perturbations can renormalize the parameters, $b_0$, $\bb$ and  $v$, but cannot open a gap. This explains why Weyl semimetal forms a stable phase \cite{murakami1}. Although the phase has yet to be experimentally observed  there are a number of proposed candidate systems, including pyrochlore iridates \cite{weyl4,witcak1}, TI multilayers \cite{weyl1,weyl2,weyl3,halasz1}, and magnetically doped TIs \cite{cho1,liu1}.   

The purpose of this Letter is to address the remarkable electromagnetic properties of Weyl semimetals. 
According to the recent theoretical work \cite{zyuzin1,son1,goswami1,grushin1}, the universal part of their EM response is described by the topological $\theta$-term,
\begin{equation}\label{s1}     
S_\theta={e^2\over 8 \pi^2}\int dt d\br \theta(\br,t) {\bm E}\cdot \bB,
\end{equation}
(using $\hbar=c=1$ units) with the `axion' angle given by 
\begin{equation}\label{th1}     
 \theta(\br,t) =2(\bb\cdot \br -b_0 t).
\end{equation}
This unusual response is a consequence of the chiral anomaly  \cite{adler1,jackiw1,nielsen1}, well known in the quantum field theory of Dirac fermions.
The physical manifestations of the $\theta$-term can be best understood from the associated equations of motion, which give rise to the following charge density and current response,
\begin{eqnarray}
\rho&=&{e^2\over 2\pi^2} \bb\cdot\bB, \label{r1} \\
\bj&=&{e^2\over 2\pi^2}(\bb\times\bE-b_0\bB). \label{r2}
\end{eqnarray}
Eq.\ (\ref{r1}) and the first term in Eq.\ (\ref{r2}) encode the anomalous Hall effect that is expected  to occur in a Weyl semimetal with broken $\cT$ \cite{weyl4,witcak1,weyl1,weyl2}. The second term in Eq.\ (\ref{r2}) describes the `chiral magnetic effect'  \cite{kenji1}, whereby a ground-state dissipationless current proportional to the applied magnetic field $\bB$ is generated in the bulk of a Weyl semimetal with broken $\cP$. 

The anomalous Hall effect is known to commonly occur in solids with broken time-reversal symmetry. In the present case of the Weyl semimetal its origin and magnitude can be understood from simple physical arguments \cite{weyl4,witcak1,weyl1,weyl2} applied to the bulk system as well as in the limit  of decoupled 2D layers \cite{grushin1}. Understanding the chiral magnetic effect (CME) in a system with non-zero energy shift $b_0$ presents a far greater challenge. The issue becomes particularly intriguing in the case of a Weyl insulator, illustrated in Fig.\ \ref{fig1}d, which will generically arise due to the exciton instability in the presence of repulsive interactions and nested Fermi surfaces. According to Ref.\ \cite{zyuzin1} CME should persist even when the chemical potential resides inside the bulk gap. At the same time, standard arguments from the band theory of solids dictate that filled bands cannot contribute to the electrical current \cite{mermin1}.  We remark that  using a different regularization scheme for the Weyl fermions Ref.\ \cite{goswami1} found that CME occurs in the semimetal but is absent in the insulator, while   Ref.\ \cite{grushin1} concluded that it only occurs when $\bb^2-b_0^2\geq m_D^2$, where $m_D$ denotes the gap magnitude.
%If this should be the case, however, then the ground state current response would exhibit an unphysical discontinuity as the gap size approaches zero. 
Semiclassical considerations  \cite{niu1} on the other hand predict a vanishing electrical current in the Weyl semimetal  but non-zero `valley current' proportional to $\bB$.        

CME, if present, could have interesting technological applications, as it constitutes a dissipationless ground state current, controllable by an external field.  Disagreements between the various field-theory predictions, however, raise important questions about the existence of CME in Weyl semimetals and insulators. The implied contradiction with one of the basic results of the band theory calls into question whether the results based on the low-energy Dirac-Weyl Hamiltonians are applicable to the real solid with electrons properly regularized on the lattice. In this Letter we undertake to resolve these questions by constructing and analyzing a lattice model of a Weyl medium. Using simple physical arguments and exact numerical diagonalization, we confirm the existence of the anomalous Hall effect as implied by Eqs.\ (\ref{r1},\ref{r2}) when $\bb\neq 0$. We find, using the same model with $b_0\neq 0$, that CME {\em does not} occur in either the  Weyl semimetal or insulator, in agreement with arguments from the band theory of solids which we review in some detail.

Our starting point is the standard model describing a 3D TI in the Bi$_2$Se$_3$ family \cite{qi_rev,fu-berg1}, regularized on a simple cubic lattice, defined by the the momentum space Hamiltonian
\begin{eqnarray}\label{h0}     
H_0(\bk)&=& 2\lambda\sigma_z(s_x\sin{k_y}-s_y\sin{k_x})+2\lambda_z\sigma_y\sin{k_z} 
\nonumber \\
&+&\sigma_xM_\bk,
\end{eqnarray}
with ${\bm \sigma}$ and $\bs$ the Pauli matrices in orbital and spin space, respectively, and $M_\bk=\epsilon-2t\sum_\alpha\cos{k_\alpha}$. For $\lambda, \lambda_z>0$ and $2t<\epsilon<6t$ the above model describes a strong topological insulator with the $Z_2$ index (1;000). In the following, we shall focus on the vicinity of the phase transition to the trivial phase that occurs at $\epsilon=6t$, via the gap closing at $\bk=0$. 

It is easy to see that Weyl semimetal emerges when we add the following perturbation to $H_0$,
\begin{equation}\label{h0}     
H_1(\bk)=b_0\sigma_y s_z+\bb\cdot(-\sigma_x s_x,\sigma_x s_y,s_z).
\end{equation}
Nonzero $b_0$ breaks $\cP$ but respects $\cT$ while $\bb$ has the opposite effect. The two symmetries are generated as follows, $\cP$: $\sigma_xH(\bk)\sigma_x=H(-\bk)$ and  $\cT$: $s_yH^*(\bk)s_y=H(-\bk)$.
\begin{figure}[t]
\includegraphics[width = 8cm]{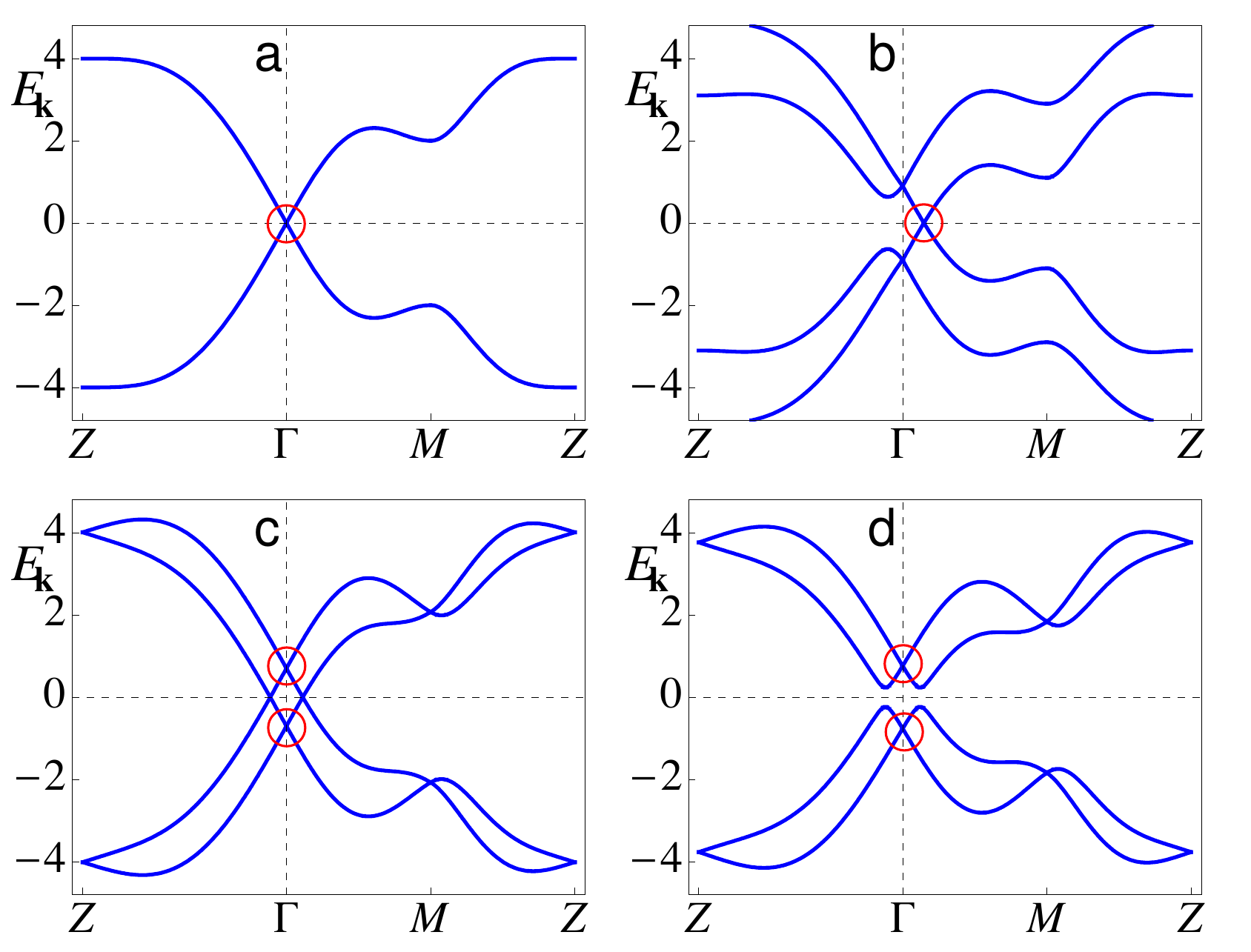}
\caption{The band structure of the Weyl semimetal lattice model, displayed along the path $\bk: \ \ (\pi,0,\pi)\to(0,0,0)\to(0,0,\pi)\to(\pi,0,\pi)$. a) Doubly degenerate 3D Dirac point when $H_1=0$ and $\epsilon=6t$. b) Momentum-shifted Weyl point for $b=0.9$ and $b_0=0$. c) Energy-shifted Weyl points for $b_z=0$ and $b_0=0.7$. d) Weyl insulator with $b_z=0$ and $b_0=0.7$ and the exciton gap modeled by taking $\epsilon=5.9t$. In all panels we take $\lambda=\lambda_z=1.0$, $t=0.5$ and the energy is measured in units of $\lambda$.  Red circles mark the location of the Dirac/Weyl points.
}\label{fig2}
\end{figure}
For simplicity and concreteness we focus on the case $\bb=b_z\hat{z}$, which yields a pair of Weyl points at $\bk=\pm (b_z/2\lambda_z)\hat{z}$. The band structure of $H=H_0+H_1$ for various cases of interest is displayed in Fig.\ \ref{fig2}.

We now address the anomalous Hall effect by directly testing Eq.\ (\ref{r1}). To this end we consider a rectangular sample of the Weyl semimetal with a base of $(L\times L)$ sites in the $x$-$y$ plane and periodic boundary conditions, infinite along the $z$-direction. The effect of the applied magnetic field is included via the standard Peierls substitution, $t\to t\exp{[2\pi i/\Phi_0\int_i^j{\bA}\cdot d{\bm l}]}$, where $\Phi_0=hc/e$ is the flux quantum, $\bA$ is the vector potential and the integral is taken along the straight line between sites $\br_i$ and $\br_j$ of the lattice. For $\bB=\hat{z}B(x,y)$ we retain the translational invariance along the $z$-direction and the Hamiltonian becomes a matrix of size $16L^2$ for each value of $k_z$. We find the eigenstates $\phi_{n,k_z}(x,y)$ of $H$ by means of exact numerical diagonalization and use these to calculate the charge density 
\begin{equation}\label{rho}
\rho(x,y)=e\sum_{n\in {\rm occ}}\sum_{k_z}|\phi_{n,k_z}(x,y)|^2.
\end{equation}
\begin{figure}[t]
\includegraphics[width = 8.5cm]{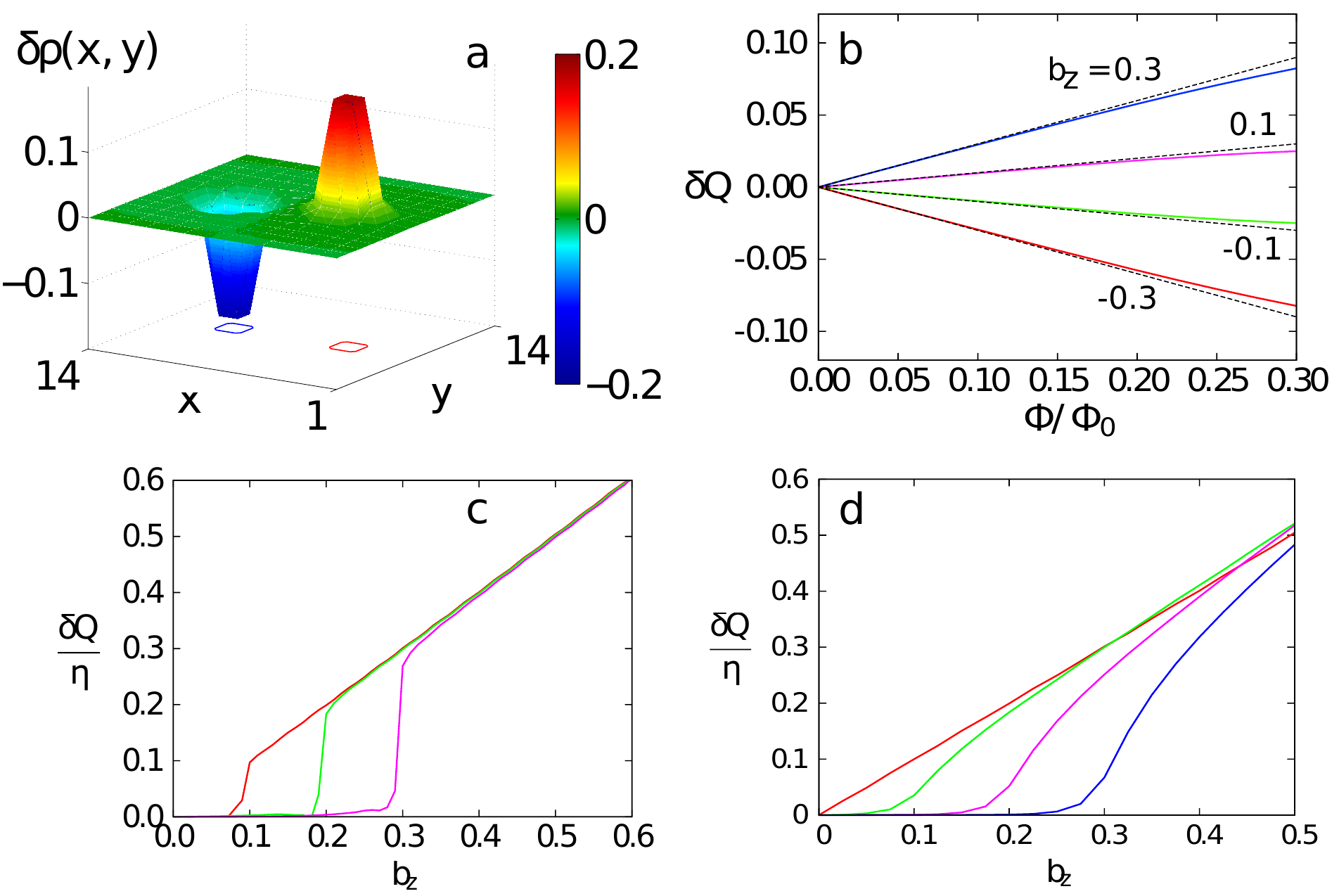}
\caption{ a) Charge density $\delta\rho(x,y)$ accumulated in the vicinity of the flux tubes $\Phi=0.01\Phi_0$ in the Weyl semimetal.
b) Total accumulated charge per layer $\delta Q$ near one of the flux tubes, in units of $e/2\pi$ for indicated values of $b_z$. Dashed lines represent the expectation based on Eq (\ref{dQ}). We use $
\lambda=\lambda_z=t=0.5$, $\epsilon=3.0$, $L=14$ and $L_z=160$ independent values of $k_z$. 
Panels c) and d) show the charge accumulations as a function of $b_z$ in the presence of non-zero Dirac mass and $b_0$. Parameters as above except $b_0= 0.1, 0.2, 0.3$ in c) and $\epsilon=3.0, 2.9, 2.8, 2.7$ for the curves in d) from left to right.
}\label{fig3}
\end{figure}
Figure \ref{fig3}a displays $\rho$ for the magnetic field configuration $B(x,y)=\Phi[\delta(x-L/4)-\delta(x+L/4)]\delta(y)$, i.e.\ two flux tubes separated by $L/2$ along the $x$ direction. In accord with Eq.\ (\ref{r1}) charge accumulates near the flux tubes, although $\rho(x,y)$ is somewhat broadened compared to $B(x,y)$. We expect the total accumulated charge per layer $\delta Q$ to be proportional to the total flux, 
\begin{equation}\label{dQ}
\delta Q= {e\over \pi}\left({b_z\over 2\lambda_z}\right) {\Phi\over\Phi_0},
\end{equation}
where we have restored the physical units.  Fig.\ \ref{fig3}b shows that this proportionality holds very accurately when the flux through an elementary plaquette is small compared to $\Phi_0$. [When the flux approaches $\Phi_0/2$ we no longer expect Eq.\ (\ref{dQ}) to hold because of the lattice effects.] We have also tested the effect of a non-zero Dirac mass, $m_D=\epsilon-6t$, and non-zero $b_0$ on the anomalous Hall effect. These terms compete with $b_z$ and for $m_D^2+b_0^2>b_z^2$ one expects the Hall effect to disappear \cite{goswami1,grushin1}. This is indeed what we observe in Figs.\ \ref{fig3}c,d. We have performed similar calculations for other field profiles $B(x,y)$ reaching identical conclusions for the anomalous Hall effect.  

We now address the chiral magnetic effect, predicted to occur when $b_0\neq 0$. We consider the same sample geometry as above, but now with uniform field $\bB=\hat{z}B$. In order to account for possible contribution of the surface states we study systems with both periodic and open boundary conditions along $x$. To find the current response we introduce a {\em uniform} vector potential $A_z$ along the $z$-direction (in addition to $A_x$ and $A_y$ required to encode the applied magnetic field). The second-quantized Hamiltonian then reads
\begin{equation}\label{ha}
{\cal H}(A_z)=\sum_{k_z}H^{\alpha\beta}(k_z-eA_z)c^\dagger_{k_z\alpha}c_{k_z\beta},
\end{equation}
where $\alpha$, $\beta$ represent all the site, orbital and spin indices. The current operator is given by 
\begin{equation}\label{curr1}
{\cal J}_z={\partial{\cal H}(A_z)\over\partial A_z}\biggl{|}_{A_z\to 0}=
-e\sum_{k_z}{\partial H^{\alpha\beta}(k_z)\over\partial k_z}c^\dagger_{k_z\alpha}c_{k_z\beta}.
\end{equation}
This leads to the current expectation value
\begin{equation}\label{curr2}
J_z=
-e\sum_{n,k_z}\left\langle\phi_{n,k_z}\left|{\partial H(k_z)\over\partial k_z}\right| \phi_{n,k_z}\right\rangle n_F[\epsilon_{n}(k_z)],
\end{equation}
where $n_F$ indicates the Fermi-Dirac distribution and $\epsilon_{n}(k_z)$ the energy eigenvalues of $H(k_z)$. We note that Eq.\ (\ref{curr2}) remains valid in the presence of the exciton condensate as long as it is treated in the standard mean field theory. 

We have evaluated $J_z$ from Eq.\ (\ref{curr2}) for various system sizes, boundary conditions, field strengths and parameter values corresponding to energy- and momentum-shifted Weyl semimetals and insulators. In all cases we found $J_z=0$ to within the numerical accuracy of our computations, typically 6-8 orders of magnitude smaller than CME expected on the basis of Eq.\ (\ref{r2}). 

For an insulator, vanishing of $J_z$  comes of course as no surprise. At $T=0$ and using the fact that $\partial_{k_z}\langle \phi_{n,k_z}|\phi_{n,k_z}\rangle=0$ one can rewrite Eq.\ (\ref{curr2}) as
\begin{equation}\label{curr3}
J_z=
-e\sum_{n\in \rm occ}\int_{\rm BZ}{dk_z\over 2\pi}{\partial \epsilon_n(k_z)\over\partial k_z},
\end{equation}
which vanishes owing to the periodicity of $\epsilon_n(k_z)$ on the Brillouin zone. More generally, for a system at non-zero temperature and when partially filled bands are present we can rewrite Eq.\ (\ref{curr2}) as 
\begin{equation}\label{curr4}
J_z=
-e\sum_{n}\int_{\rm BZ}{dk_z\over 2\pi}{\partial \epsilon_n(k_z)\over\partial k_z} n_F[\epsilon_{n}(k_z)],
\end{equation}
where the sum over $n$ extends over all bands. 
By transforming the $k_z$-integral in Eq.\ (\ref{curr4}) into an integral over the energy it is easy to see that it identically vanishes for any continuous energy dispersion $\epsilon_{n}(k_z)$ that is periodic on the Brillouin zone and for any distribution function that only depends on energy. This reflects the well-known fact that one must establish a non-equilibrium distribution of electrons to drive current in a metal, e.g.\ by applying an electric field.   Given these arguments we conclude that, as a matter of principle, CME cannot occur in a crystalline solid, at least when interactions are unimportant and the description within the independent electron approximation remains valid.

There are several notable cases when filled bands {\em do} contribute currents. A superconductor can be thought of as an insulator for Bogoliubov quasiparticles and yet it supports a supercurrent. This occurs because Bogoliubov quasiparticles, being coherent superpositions of electrons and holes, do not carry a definite charge and consequently the current cannot be expressed through Eq.\ (\ref{curr2}). In quantum Hall insulators non-zero $\sigma_{xy}$ also  implies non-vanishing current. In the standard Hall bar geometry, used in transport measurements, it is well known that the physical current is carried by the gapless edge modes, not through the gapped bulk. In the Thouless charge pump geometry the current indeed flows through the insulating bulk but this requires a time-dependent Hamiltonian (the magnetic flux through the cylinder is time dependent). Our considerations leading to Eq.\ (\ref{curr2}) are only valid for time-independent Hamiltonians. Finally, there are known cases \cite{esteve1} when the transition from Eq.\ (\ref{curr2}) to  (\ref{curr3}) fails because the Hamiltonian is not self-adjoint on the space of functions that includes derivatives of $\phi$. This can happen when the Hamiltonian is a differential operator but in our case $H(k_z)$ is a finite-size hermitian matrix with a smooth dependence on $k_z$, which precludes any such exotic possibility. In any case, our numerical calculations addressed directly  Eq.\ (\ref{curr2}) so self-adjointness cannot possibly be an issue. 

\begin{figure}[t]
\includegraphics[width = 8.5cm]{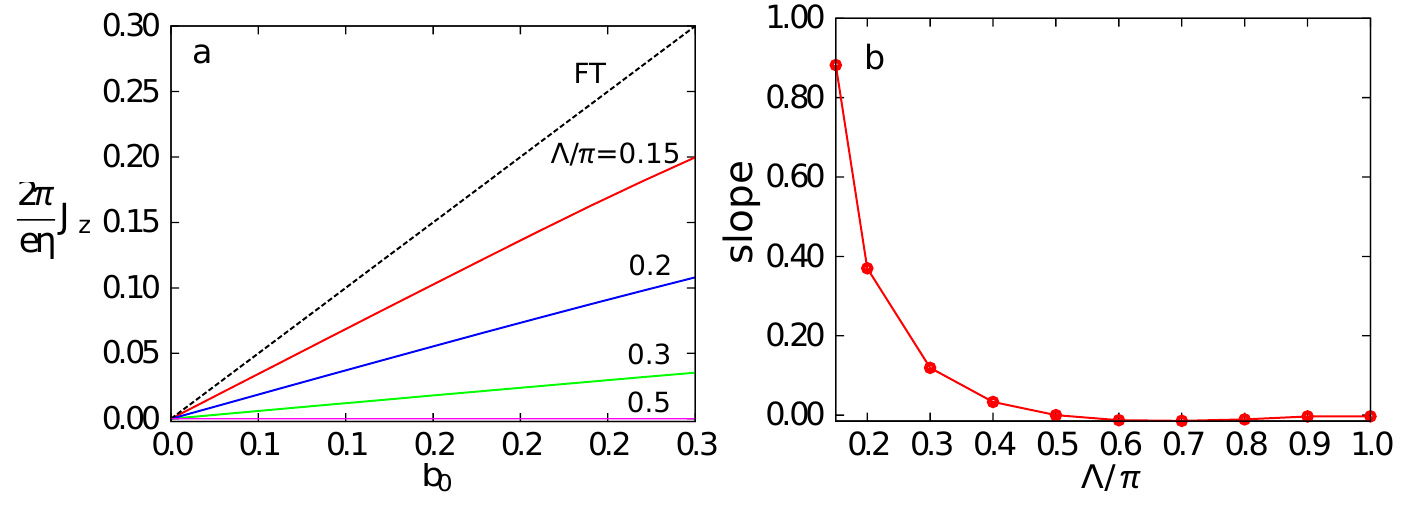}
\caption{a) Chiral current $J_z$ as a function of energy offset $b_0$ for various values of the momentum cutoff $\Lambda$. The dashed line indicates the field theory prediction Eq.\ (\ref{r2}). b) The slope $dJ_z/db_0$ in units of $e\eta/2\pi$ as a function of cutoff $\Lambda$. Slope 1.0 is expected on the basis of Eq.\ (\ref{r2}).
}\label{fig4}
\end{figure}
Our considerations conclusively establish that anomalous Hall effect \cite{weyl4,witcak1,weyl1,weyl2}, quantitatively  consistent with the prediction of the low-energy continuum theory \cite{zyuzin1,son1,goswami1,grushin1}, occurs in realistic Weyl semimetals defined on the  lattice. The chiral magnetic effect \cite{kenji1}, implied by the same considerations via Eq.\ (\ref{th1}), however runs afoul of the basic results of the band theory and is found to be absent. Within the low-energy continuum theory the form of  the axion angle given  Eq.\ (\ref{th1}) can be expected on the basis of Lorenz invariance. In the real solid  this symmetry is broken at the lattice scale so there is no fundamental reason why this form should hold beyond the low-energy approximation. To test this hypothesis we have evaluated  current $J_z$ from Eq.\ (\ref{curr2}) with a momentum cutoff, i.e.\ limiting $|k_z|<\Lambda$ in the sum. As shown in Fig.\ (\ref{fig4}) when $\Lambda\ll\pi$ one indeed obtains CME with a magnitude consistent with Eq.\ (\ref{r2}). The current however rapidly vanishes as the cutoff approaches the extent of the full Brillouin zone.  We remark that imposing such a cutoff has no significant effect on the Hall effect calculation as long as $\Lambda>|b_z/2\lambda_z|$ because the entire Hall response comes from this region of the momentum space \cite{weyl4,witcak1,weyl1,weyl2}. These considerations thus explain the difference between the low-energy and lattice descriptions of Weyl semimetals.

In closing, we note that if correct, the time dependence of the axion angle implied by  Eq.\ (\ref{th1}) would engender some peculiar consequences. One of them follows from the Witten effect \cite{witten1,rosenberg1} whereby a unit magnetic monopole inserted into the axion medium carries a polarization charge $\delta Q=-e(n+\theta/2\pi)$ with $n$ integer. In the Weyl semimetal with a nonzero energy shift $b_0$ this charge $\delta Q$ would grow linearly with time according to  Eq.\ (\ref{th1}). Although such `quantum time crystal'  behavior has been conjectured to arise in certain interacting systems \cite{wilczek_time} it is not clear by what mechanism it would occur in the ground state of a non-interacting semimetal.  Our findings indeed confirm the  absence of this behavior in the Weyl semimetal described by a natural lattice Hamiltonian. It remains an open question whether this fascinating phenomenon can be realized in another quantum system. Another interesting problem which we leave for future investigation is finding the regularization scheme for the low-energy theory that would match the results of our lattice calculation. 

The authors are indebted to I. Affleck, A.A. Burkov, M.P.A. Fisher, A. Grushin and I.F. Herbut for illuminating discussions and correspondence. The work was supported by NSERC and CIfAR. 

%...

\end{document}